\documentclass{aa}
\usepackage{graphicx}
\usepackage{amsmath}

\input{tcilatex}

\begin{document}
\thesaurus{12(12.07.1; 08.02.1; 08.16.2)}

\title{Caustics in special multiple lenses}
\author{Valerio Bozza\thanks{%
E-mail valboz@physics.unisa.it}} \institute{Dipartimento di
Scienze Fisiche E.R. Caianiello,\\
 Universit\`{a} di Salerno, I-84081 Baronissi, Salerno, Italy.}
\date{Received / Accepted }
\maketitle

\begin{abstract}
Despite its mathematical complexity, the multiple gravitational
lens can be studied in detail in every situation where a
perturbative approach is possible. In this paper, we examine the
caustics of a system with a lens very far from the others with
respect to their Einstein radii, and a system where mutual
distances between lenses are small compared to the Einstein radius
of the total mass. Finally we review the case of a planetary
system adding some new information (area of caustics, duality and
higher order terms). %
\end{abstract}

\section{Introduction}

The investigation of a gravitational lens composed by a discrete
set of point-like masses was opened by Schneider \& Wei{\ss} in
1986. They considered a binary lens, deriving analytical
expressions for caustics in the case of lenses with the same mass.
Erdl \& Schneider (1993) extended their analysis to an arbitrary
mass ratio, confirming the presence of three possible topologies
in the caustic structure.

Besides these two fundamental papers, many other contributions
have been given to this field by several authors studying
particular aspects of this matter. It has often been considered
the relation between the binary lens and the Chang--Refsdal (1979;
1984) model describing the behaviour of a point-like lens immersed
in a uniform gravitational field (Schneider \& Wei{\ss} 1986;
Dominik 1999). In the special case of very asymmetric lens, as
that of a planet orbiting a star, the Chang--Refsdal approximation
can give very reliable results (Gould \& Loeb 1992; Gaudi \& Gould
1997). In the search for good approximate models, also the
quadrupole lens was given a role in close binary systems (Dominik
1999).

However, while the picture for the binary lens is somewhat well
defined at present day, the same cannot be said about general
multiple lenses. The attempts to enlighten this kind of lenses
have really been very few. This is because of the mathematical
complexity of the lens equation. Yet, multiple lenses can play an
important role in several cases and therefore are not just
mathematical curiosities. Planetary systems are likely to be
composed by a star with more than one planet, so it is important
to study them in their full complexity in order to interpret
planetary microlensing events correctly, especially those with
high magnification (Wambsganss 1997; Gaudi, Naber \& Sackett 1998;
Bozza 1999). N-body systems made of similar masses are hard to be
found as gravitationally stable systems, yet the projected
distances between far stable subsystems could be compatible with
the appearance of ``interaction'' in the gravitational lensing
behaviour; some considerations about microlensing light curves by
multiple lenses have been done by Rhie (1996).

Another context where multiple lensing is surely of great interest
is cosmological lensing by rich galactic clusters: as a first
approximation, all lensing galaxies can be considered as point
masses, then the multiple Schwarzschild lens constitutes a good
starting point for the analysis of caustics and images produced by
such systems.

The aim of this work is to give a systematic investigation of the
caustics of multiple lenses whenever they can be referred to
well--defined situations to be employed as starting points in
perturbative expansions. The three cases where this happens have
been pointed out by Dominik (1999) and are resumed here.

In Sect. 2 the lens equation for multiple lenses and the basic
definitions are given; Sect. 3 deals with the case of a lensing
mass that is very far from the others; in Sect. 4 a system formed
by near (with respect to the Einstein radius of the total mass)
lenses is studied; Sect. 5 reconsiders the case of a multiple
planetary system, already examined in a former work (Bozza 1999);
Sect. 6 contains the conclusions.

\section{Basics of multiple lensing}

Let
\begin{equation}
R_{\mathrm{E}}^{\mathrm{0}}=\sqrt{\frac{4GM_{\mathrm{0} }}{c^{2}}
\frac{D_{\mathrm{LS}}D_{\mathrm{OL}}}{D_{\mathrm{OS}}}}
\end{equation}
be the Einstein radius of a mass $M_{\mathrm{0}}$ placed at
distance $D_{\mathrm{OL}}$ from the observer and at distance
$D_{\mathrm{LS}}$ from a source having distance
$D_{\mathrm{OS}}=D_{\mathrm{OL}}+D_{\mathrm{LS}}$ from the
observer. The coordinates in the lens plane normalized to
$R_{\mathrm{E}}^{\mathrm{0}}$ will be denoted by $\mathbf{x}=\left(%
x_{1};x_{2}\right) $, while the coordinates in the source plane
normalized to
$R_{\mathrm{E}}^{\mathrm{0}}\frac{D_{\mathrm{OS}}}{D_{\mathrm{OL}}}$
are $\mathbf{y}=\left( y_{1};y_{2}\right) $. All masses are meant
to be measured in units of $M_{\mathrm{0}}$ which is a typical
mass of the problem we are considering. For example it can be the
solar mass if the lensing objects are stars, or a typical galaxy
mass if we are considering extragalactic lensing.

According to the standard theory of gravitational lensing, the
deviation induced by a system formed by n masses $m_{1}, ...,
m_{n}$ whose projections in the lens plane are at positions
$\mathbf{x}_{1}, ..., \mathbf{x}_{n}$ is the sum of the deviations
produced by the single objects. So the lens equation reads:
\begin{equation}
\mathbf{y}=\mathbf{x}-\sum\limits_{i=1}^{n}\frac{m_{i}\left( \mathbf{x}-%
\mathbf{x}_{i}\right) }{\left| \mathbf{x}-\mathbf{x}_{i}\right| ^{2}%
} \label{General lens equation}
\end{equation}

Given a source position $\mathbf{y}$, the $\mathbf{x}$'s solving
this equation are called images. Their properties can be studied
through the Jacobian matrix of the lens application. In
particular, the determinant of this matrix is:
\begin{multline}
\det J=1-\left[\sum\limits_{i=1}^{n}\frac{m_{i}\left( \Delta
_{i1}^{2}-\Delta _{i2}^{2}\right) }{\left( \Delta _{i1}^{2}+\Delta
_{i2}^{2}\right) ^{2}}\right] ^{2}+  \label{General Jacobian}\\
-4\left[\sum\limits_{i=1}^{n}\frac{m_{i}\Delta _{i1}\Delta
_{i2}}{\left( \Delta _{i1}^{2}+\Delta _{i2}^{2}\right)
^{2}}\right] ^{2}
\end{multline}
where $\mathbf{\Delta }_{i}=\left( \Delta _{i1};\Delta _{i2}\right) =%
\mathbf{x}-\mathbf{x}_{i}$.

The magnification of an image is given by the absolute value of
the inverse of $\det J$ calculated at the image position. Whenever
the Jacobian determinant vanishes, an image at that point would be
(in ray optics) infinitely amplified. These points are thus said
to be critical and their corresponding points in the source plane,
found by use of Eq. (\ref{General lens equation}), constitute the
caustics.

Caustics appear as closed cusped curves and split the source plane
in several domains. The number of images corresponding to a single
source only depends on the domain where the source is located. In
neighboring domains the number of images always differs by two
units. The caustics of a given lens model provide very useful
hints for the description of the general properties of the system
so that their study is essential to understand the behaviour of a
lens.

To find the critical curves and then the caustics of a multiple
lens, one has to solve the equation $\det J=0$, which is very
complicated. However, if it is possible to single out some
parameters which are very large or small with respect to the
others, a perturbative approach can be tried (Bozza 1999). The
situations we are to explore in this paper always refer to the
single Schwarzschild lens as the zero order solution. The critical
curve of a lens with mass $m$ placed at the origin is a circle
with radius $\sqrt{m}$, while the caustic is reduced to the origin
itself: $\mathbf{y}=0$.

\section{Single lens perturbed by far masses}

Let's start considering an isolated point-lens. The presence of
other masses that are very far from the first one (with respect to
all Einstein radii) warps the circular critical curve, giving rise
to an extended caustic.

Let's consider a mass $m_{1}$ placed at the origin of the lens
plane. The other masses $m_{2}, ..., m_{n}$ are at the positions
$\mathbf{x}_{2}, ..., \mathbf{x}_{n}$. In polar coordinates, we
put $\mathbf{x}_{i}=\left( \rho_{i} \cos \varphi_{i};\rho_{i} \sin
\varphi_{i} \right)$ and the generic coordinate in the lens plane
is $\mathbf{x}=\left( r \cos \vartheta;r \sin \vartheta \right)$.
The hypothesis we start from is that $\rho_{i}\gg \sqrt{m_{j}}$
for each $i$ and $j$. This allows us to expand the Jacobian
determinant (\ref{General Jacobian}) in series of inverse powers
of $\rho_{i}$. If we stop at the fourth order we have:
\begin{multline}
\det J=1-\frac{m_{1}^{2}}{r^{4}}+\\%
-\frac{2 m_{1}}{r^{2}}
\sum\limits_{i=2}^{n} \frac{ m_{i} \cos \left( 2 \vartheta -2
\varphi_{i}\right)}{
\rho_{i}^{2}}+\\%
-\frac{4 m_{1}}{r}\sum\limits_{i=2}^{n} \frac{m_{i} \cos \left( 3
\vartheta -3 \varphi_{i}\right)}{ \rho_{i}^{3}}+\\%
 -6 m_{1}\sum\limits_{i=2}^{n} \frac{ m_{i} \cos \left( 4 \vartheta -4
\varphi_{i}\right)}{\rho_{i}^{4}}
-\sum\limits_{i,j=2}^{n}\frac{m_{i} m_{j} \cos \left( 2
\varphi_{i}-2 \varphi_{j}\right)}{\rho_{i}^{2}
\rho_{j}^{2}}+\\%
+o\left( \frac{1}{\rho_{i}^{4}}\right)%
\label{Far Jacobian}
\end{multline}

We see that a linear superposition principle is valid at the
second and the third order. In the fourth order there are
``interaction'' terms that make the calculations more difficult.

The zero order solution is simply the Einstein radius
$r=\sqrt{m_1}$. We build the general solution as an infinite
series in $1/\rho_i$:
\begin{equation}
r=\sqrt{m_1}\left( 1+ \varepsilon_1+
\varepsilon_2+\varepsilon_3+...\right)%
\label{r expansion}
\end{equation}
where $\varepsilon_j \sim 1/\rho_i^j$. Substituting this
expression in Eq. (\ref{Far Jacobian}) and expanding again, we can
easily solve the equation of critical curves at each order to find
the perturbations. The first three orders are:
\begin{equation}
\begin{array}{l}
\varepsilon_1=0 \\%
\varepsilon_2=\sum\limits_{i=2}^{n}\frac{m_{i} \cos
\left(2\vartheta-2\varphi_i \right)}{2\rho_i^2}\\%
\varepsilon_3=\sum\limits_{i=2}^{n}\frac{\sqrt{m_1} m_{i} \cos
\left(3\vartheta-3\varphi_i \right)}{\rho_i^3}
\end{array}
\end{equation}

Recalling that the shear produced by one mass in a Chang--Refsdal
approximation is $\gamma_i=\frac{m_i}{\rho_i^2}$, the second order
can be easily recognized as the sum of the Chang--Refsdal
expansions for each mass to the first order in the $\gamma_i$.
Actually, the full second order can be identified with a
Chang--Refsdal lens expanded to the first order in this total
shear:
\begin{equation}
\gamma=\sqrt{\left[ \sum\limits_{i=2}^n \frac{m_i \sin \left(
2\varphi_i \right)}{\rho_i^2}\right]^2+\left[ \sum\limits_{i=2}^n
\frac{m_i \cos \left( 2\varphi_i \right)}{\rho_i^2}\right]^2}%
\label{Wide caustic shear}
\end{equation}
and oriented along the direction at the angle:
\begin{equation}
\varphi=\frac{1}{2}\arctan \left[ \frac{\sum\limits_{i=2}^n
\frac{m_i \sin \left( 2\varphi_i
\right)}{\rho_i^2}}{\sum\limits_{i=2}^n \frac{m_i \cos \left(
2\varphi_i \right)}{\rho_i^2}} \right]%
\label{Wide caustic angle}
\end{equation}

So, the Chang--Refsdal lens provides a first approximation for the
critical curves of a far lens not only in the binary case (Dominik
1999) but even for an arbitrary number of lenses, as already
stated in (Chang \& Refsdal 1979).

Now we exploit the perturbative results just shown to find the
caustic formed in this case. It suffices to apply the lens
equation (\ref{General lens equation}) and expand it to the third
order again:
\begin{mathletters}
\begin{eqnarray}
& & y_1\left( \vartheta \right) = \sum\limits_{i=2}^n \frac{m_i
\cos \left( \varphi_i
\right)}{\rho_i}+ \nonumber \\%
& & +\sqrt{m_1}\sum\limits_{i=2}^n \frac{m_i}{\rho_i^2}\left[\cos
\vartheta \cos \left(2\vartheta-2 \varphi_i \right)+\cos
\left(\vartheta-2 \varphi_i \right) \right]+\nonumber  \\%
& & +m_1\sum\limits_{i=2}^n \frac{m_i}{\rho_i^2} \left[2\cos
\vartheta  \cos \left(3\vartheta-3 \varphi_i \right)+ \cos
\left(2\vartheta-3 \varphi_i \right)\right]  \nonumber\\ & & \\%
& & y_2\left( \vartheta \right) = \sum\limits_{i=2}^n \frac{m_i
\sin \left( \varphi_i
\right)}{\rho_i}+\nonumber  \\%
& & +\sqrt{m_1}\sum\limits_{i=2}^n\frac{m_i}{\rho_i^2} \left[\sin
\vartheta \cos \left(2\vartheta-2 \varphi_i \right)-\sin
\left(\vartheta-2 \varphi_i \right) \right]+\nonumber \\%
& & +m_1\sum\limits_{i=2}^n \frac{m_i}{\rho_i^2} \left[2\sin
\vartheta \cos \left(3\vartheta-3 \varphi_i \right)- \sin
\left(2\vartheta-3 \varphi_i \right)\right] \nonumber \\ & &
\end{eqnarray}
\end{mathletters}

The first term is independent of $\vartheta$ and only fixes the
position of the caustic which is displaced towards the other
masses; the two second order terms describe the shape of the
caustic reproducing the Chang--Refsdal limit; the third order
terms correct this approximation.

The Chang--Refsdal model describes the gravitational lensing by a
mass embedded in a uniform gravitational field. For $\gamma>1$,
the caustics of this model are diamond--shaped curves with four
cusps which are independent on the versus of the gravitational
field. The approximation of uniform field is reasonable in our
case. However, it breaks down quite soon when the distances are
not so high. It happens that our lens begins to feel the non
uniformity of the field and the caustics lose their symmetry,
elongating in the direction of the other masses. The third order
describes this elongation very well, considerably extending the
range of applicability of the perturbative results.

The breakdown of this perturbative expansion rises when the
Einstein ring of the first mass is too close to the critical
curves produced by the other masses. A particular case is obtained
when several small masses are very close each other. The critical
curve generated by such a subsystem has a radius of the order of
the square root of its total mass. So even if the condition
$\rho_{i}\gg \sqrt{m_{j}}$ is satisfied for each single mass, the
distance of the first mass to the subsystem could be smaller than
the radius of its total critical curve, causing the failure of the
perturbative hypothesis.

\begin{figure}
 \resizebox{\hsize}{!}{\includegraphics{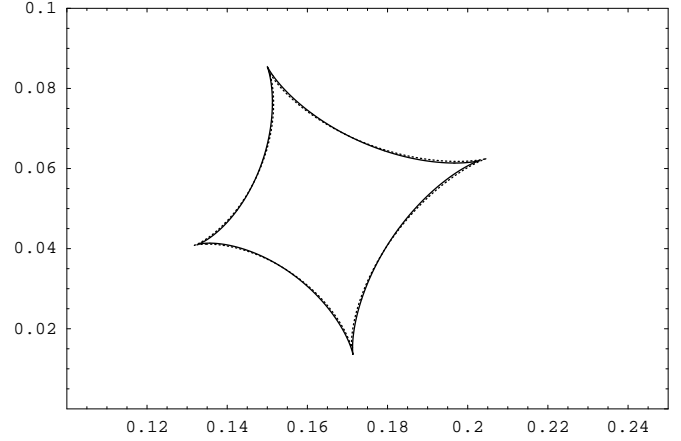}}
 \caption{The caustic of a lens perturbed by two other bodies. %
 The masses are $m_1=0.25$, $m_2=0.25$, $m_3=0.5$. %
 The positions of $m_2$ and $m_3$ are %
 $\rho_2=4$, $\varphi_2=0$ and $\rho_3=\sqrt{20}$, $\varphi_3=\arctan 0.5$ respectively. %
 The solid line is the perturbative caustic to be compared with the dashed line that is %
 the exact one, almost completely hidden behind the solid line.}
 \label{Fig wide ternary caustic}
\end{figure}

\begin{figure*}
 \resizebox{\hsize}{!}{\includegraphics{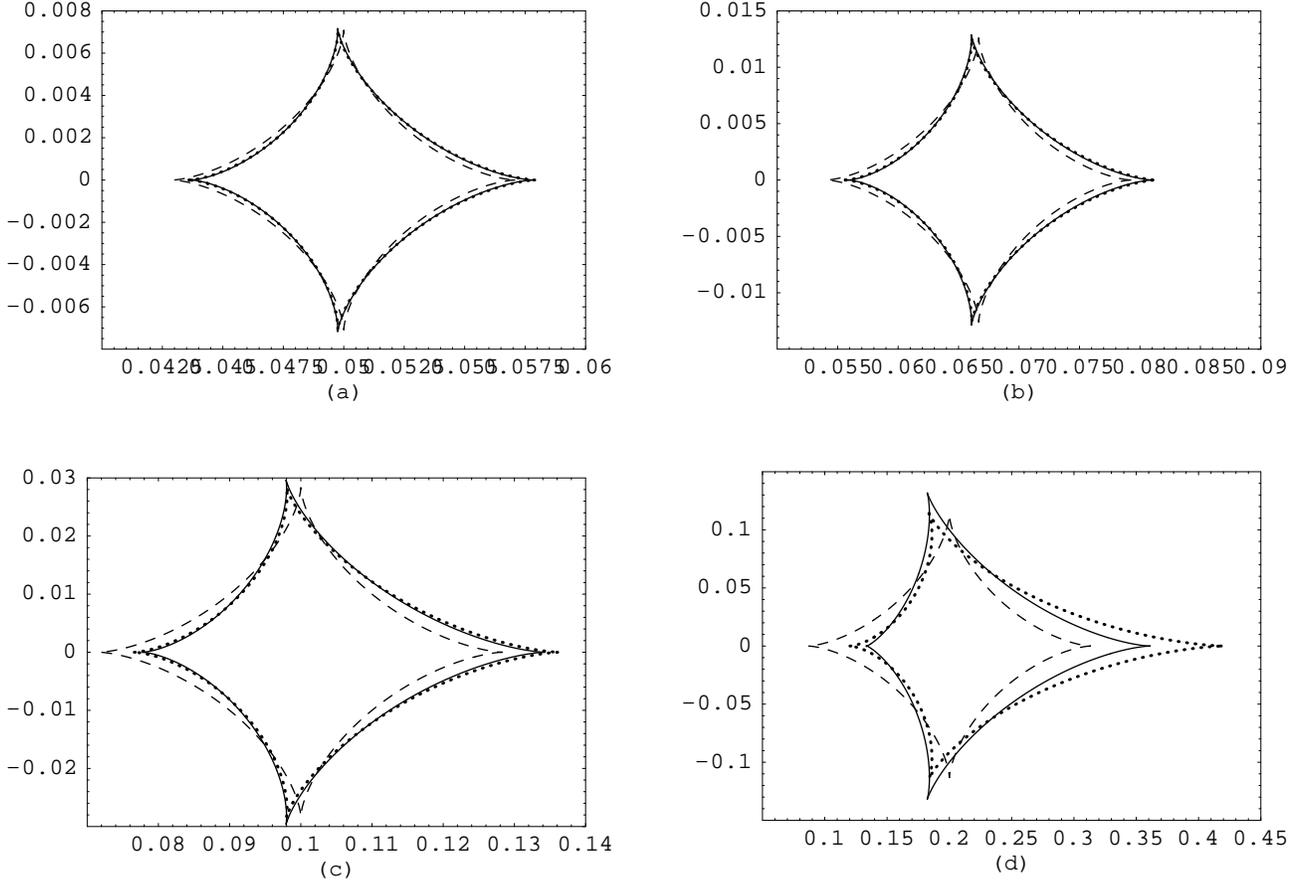}}
 \caption{Caustic of a lens with a far companion of the same mass ($0.5$), %
 along the direction $\varphi_2=0$. The dotted curves are the exact %
 caustics, the dashed curves are the second order perturbative ones %
 (Chang--Refsdal) and the continuous lines are the third order ones. %
 (a) $\rho_2=10$, (b) $\rho_2=7.5$, (c) $\rho_2=5$, (d) $\rho_2=2.5$.}
 \label{Fig wide binaries caustics}
\end{figure*}

In Fig. \ref{Fig wide ternary caustic}, we see an example of the
power of the perturbative approach in reproducing the numerically
found exact results with great accuracy not taking care of the
number of masses.

Fig. \ref{Fig wide binaries caustics} shows an interesting
comparison among the different orders approximations and the exact
caustic for different distances between the two lenses. We see
that even at separations about $10\sqrt{2}$ times the Einstein
radius of each lens, the second order approximation
(Chang--Refsdal) seems quite poor, while the third order caustic
is practically coincident with the exact one up to five Einstein
radii. As previously said the inadequacy of the Chang--Refsdal
approximation is in its impossibility in giving account of the
asymmetry of the caustic, which is already evident at quite far
separations. This justifies us in considering the third order that
is the lowest order reproducing this asymmetry.

Now let's turn to the cusps of these caustics. These points are
characterized by the vanishing of the tangent vector:
\begin{equation}
\left\{
\begin{array}{l}
y_{1}^{\prime }\left( \vartheta \right) =0 \\ y_{2}^{\prime
}\left( \vartheta \right) =0
\end{array}
\right. %
\label{General cusp equation}
\end{equation}

At the second order we have a Chang--Refsdal caustic rotated at an
angle $\varphi$ given by Eq. (\ref{Wide caustic angle}) from the
$y_1$--axis. Without loss of generality we can put this angle to
zero. Then it is well known (Dominik 1999) that the curve is
symmetric for reflections on two axes passing through the center
of the caustic and has four cusps at the intersections with these
axes, corresponding to $\vartheta=0,\frac{\pi}{2}, \pi,
\frac{3\pi}{2}$.

At the third order, summing and subtracting the two Eqs.
(\ref{General cusp equation}), we get:
\begin{equation}
\left\{
\begin{array}{l}
\left(\cos \vartheta+\sin \vartheta \right)
F\left(\vartheta\right)=0
\\ \left(\cos \vartheta-\sin \vartheta \right) F\left(\vartheta\right)=0
\end{array}
\right.
\end{equation}
with
\begin{equation}
F\left(\vartheta\right)=\sum\limits_{i=2}^{n}m_i\left[\frac{3
\cos\left(2\vartheta-2\varphi_i\right)}{\rho_i^2}+\frac{8\sqrt{m_1}
\cos\left(3\vartheta-3\varphi_i\right) }{\rho_i^3}\right]
\end{equation}

The two Eqs. can be simultaneously satisfied only if
$F\left(\vartheta\right)$ vanishes. For a binary system, this
equation has six roots:
\begin{equation}
\begin{array}{l}
\vartheta=0\\ %
\vartheta=\pi\\%
\vartheta=\arccos \frac{-3\rho_2+\sqrt{256m_1+9\rho_2^2}}{32\sqrt{m_1}}\\ %
\vartheta=-\arccos \frac{-3\rho_2+\sqrt{256m_1+9\rho_2^2}}{32\sqrt{m_1}}\\ %
\vartheta=\arccos \frac{-3\rho_2-\sqrt{256m_1+9\rho_2^2}}{32\sqrt{m_1}}\\ %
\vartheta=-\arccos \frac{-3\rho_2-\sqrt{256m_1+9\rho_2^2}}{32\sqrt{m_1}}\\ %
\end{array}
\end{equation}

\begin{figure}
 \resizebox{\hsize}{!}{\includegraphics{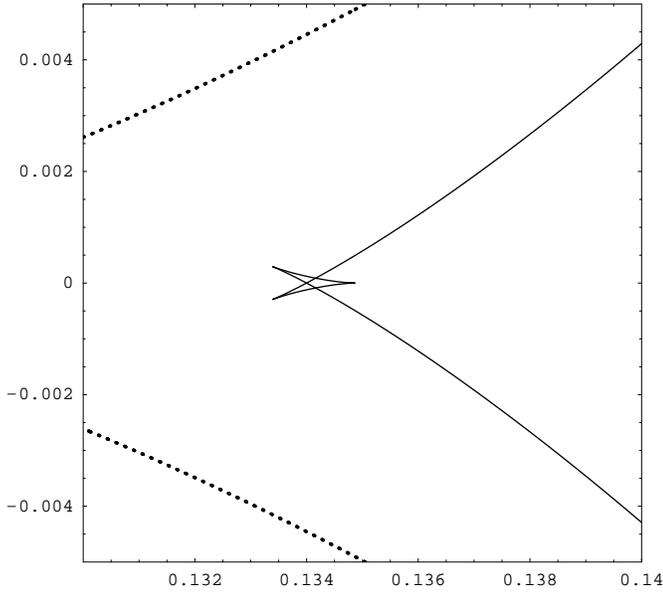}}
 \caption{A particular of Fig. \ref{Fig wide binaries caustics}d showing %
 the two additional cusps yielded by the perturbative expansion.}
 \label{Fig wide binary particular}
\end{figure}

The first two are the cusps along the $y_1$--axis already present
at the second order. The second two are the modification of the
other two Chang--Refsdal cusps (in fact, when $\rho_2$ tends to
infinity, they approach $\pm \frac{\pi}{2}$). The last two cusps
are imaginary for $\rho_2>4\sqrt{m_1}$, but become real for lower
distances, giving rise to a ``butterfly'' geometry around
$\vartheta=\pi$ (Fig. \ref{Fig wide binary particular}). For
binary systems, this is not what happens in exact results which
always yield four cusps; yet, with more than two lenses, an
increment in the number of cusps is effectively present when the
separation between the lenses is not very high.

Having at our disposal good approximate expressions for the
caustics of multiple lenses, an interesting quantity to compute is
the area covered by these curves in the source plane. We do this
calculation for an arbitrary multiple lens at the second order and
for a binary system at the third order.

For the Chang--Refsdal approximation, again we assume that the
angle $\varphi$ defined in Eq. (\ref{Wide caustic angle}) is zero.
Then the caustic is oriented along the $y_1$--axis and its area is
given by:
\begin{equation}
A=2\int\limits_{y_1^{\mathrm{min}}}^{y_1^{\mathrm{max}}} y_2
\mathrm{d}y_1
\end{equation}

We have previously expressed $y_2$ and $y_1$ as functions of
$\vartheta$. Considering that, when $\vartheta=0$, $y_1$ is
$y_1^{\mathrm{max}}$ and, when $\vartheta=\pi$, $y_1$ is
$y_1^{\mathrm{min}}$, also considering that $y_2$ is negative for
$0<\vartheta<\pi$ and positive for $\pi<\vartheta<2\pi$, the area
can be calculated by the following integral:
\begin{equation}
A=\int\limits_{0}^{2\pi} y_2\left( \vartheta \right)
\frac{\mathrm{d}y_1}{\mathrm{d}\vartheta}\mathrm{d}\vartheta %
\label{Area formula}
\end{equation}

The result is:
\begin{equation}
A=\frac{3}{2}\pi m_1\gamma^2
\end{equation}

Remembering the expression (\ref{Wide caustic shear}) for $\gamma$
in the physical system we are analyzing, we see that the main
dependences of the area of the caustic on its parameters are
quadratic in the masses and inverse fourth power type in the
distances.

For the third order calculation, a direction of symmetry can no
longer be determined unless we limit ourselves to the binary
system. The procedure just exposed can be followed for $\rho_2>4
\sqrt{m_1}$, when the geometry of the caustic remains unaltered.
Then, we obtain a correction to the previous result:
\begin{equation}
\Delta A=4\pi \frac{m_1^2 m_2^2 }{\rho_2^6}
\end{equation}
So, not only the third order stretches the caustic but also adds a
positive contribution to its area.

\section{Close multiple lenses}

Now we consider a set of $n$ point-like lenses whose mutual
distances are small compared to the total Einstein radius of the
system. The starting point for our analysis is the Schwarzschild
lens we would obtain if the total mass $M$ were concentrated at
the barycentre. The separations between the masses modulate the
deviations from the Schwarzschild lens and then constitute the
perturbative parameters in our expansion.

Now there is no privileged mass, so the most reasonable choice of
the coordinates origin would be the centre of mass. We derive our
results in the general case but we will choose the centre of mass
system for some particular considerations.

Besides the main critical curve, which is a slight modification of
the Einstein ring of the barycentral lens, some secondary critical
curves are present near the centre of mass. They give rise to very
small and far caustics which must be treated separately.

\subsection{Main critical curve and central caustic}

We consider our masses $m_{1}, ..., m_{n}$ placed at the positions
$\mathbf{x}_{1}, ..., \mathbf{x}_{n}$. In polar coordinates, we
put $\mathbf{x}_{i}=\left( \rho_{i} \cos \varphi_{i};\rho_{i} \sin
\varphi_{i} \right)$ and the generic coordinate in the lens plane
is $\mathbf{x}=\left( r \cos \vartheta;r \sin \vartheta \right)$.
Now we assume that $\rho_{i}\ll \sqrt{M}$ for each $i$. We carry
the expansion of the Eq. (\ref{General Jacobian}) in series of
powers of $\rho_{i}$ up to the second order:
\begin{multline}
\det J=1-\frac{M^{2}}{r^{4}}-\frac{4M}{r^5}\sum\limits_{i=1}^{n}
\rho_{i} m_i \cos \left( \vartheta -\varphi_{i}\right)+\\%
-\frac{2M}{r^6}\left[ 3 \sum\limits_{i=1}^{n} m_{i} M \rho_{i}^2
\cos \left( 2 \vartheta -2
\varphi_{i}\right)+\right. \\%
\left.+2\sum\limits_{i,j=1}^n m_{i}m_{j}\rho_{i}\rho_{j}
\cos \left(\varphi_{i}-\varphi_{j}\right)\right]%
\label{Close Jacobian}
\end{multline}

The zero order terms give the Einstein ring for a point mass $M$
at the origin. We see that interference terms arise at the second
order and this is a good reason to stop our expansion to avoid
long calculations at higher orders.

We can consider an expression analogue to Eq. (\ref{r expansion})
for the radial coordinate $r$:
\begin{equation}
r=\sqrt{M}\left( 1+ \varepsilon_1+
\varepsilon_2+...\right)%
\end{equation}
where now $\varepsilon_j \sim \rho_i^j$. Substituting in Eq.
(\ref{Close Jacobian}) and expanding again, we can solve for the
$\varepsilon_j$ at each order:
\begin{equation}
\begin{array}{l}
\varepsilon_1=\frac{1}{M^{3/2}}\sum\limits_{i=1}^{n} m_{i}
\rho_{i}
\cos \left( \vartheta-\varphi_{i}\right)  \\%
\varepsilon_2=\frac{1}{4 M^3}\left\{ 6\sum\limits_{i=1}^{n}M m_{i}
\rho_{i}^2 \cos
\left(2\vartheta-2\varphi_i \right) +\right. \\%
\left. - \sum\limits_{i,j=1}^n m_{i} m_{j}\rho_{i} \rho_{j} \left[
5\cos \left( 2\vartheta-\varphi_i - \varphi_j \right) + \cos
\left( \varphi_{i}- \varphi_{j} \right) \right] \right\}
\end{array}
\end{equation}

Now that the perturbations to the critical curve are known (to the
second order), we can find the caustic by use of the lens
equation:
\begin{mathletters}
\begin{eqnarray}
& & y_1\left( \vartheta \right) = \frac{1}{M}\sum\limits_{i=1}^n
m_i \rho_i \cos \varphi_i +\frac{1}{2M^{5/2}}
\sum\limits_{i=1,j\neq i}^n m_i m_j \cdot \nonumber \\%
& & \cdot \left\{ 3 \rho_i^2 \cos \left( \vartheta -2 \varphi_i
\right) +\rho_i^2 \cos
\left( 3 \vartheta -2 \varphi_i \right)+ \right. \nonumber \\%
& &\left. -\rho_i \rho_j \left[ 3 \cos \left( \vartheta -
\varphi_i -\varphi_j \right) +\cos \left( 3 \vartheta - \varphi_i
-\varphi_j \right) \right] \right\} \\%
& &y_2\left( \vartheta \right) = \frac{1}{M}\sum\limits_{i=1}^n
m_i \rho_i \sin \varphi_i - \frac{1}{2M^{5/2}}
\sum\limits_{i=1,j\neq i}^n m_i m_j \cdot \nonumber \\%
& & \cdot \left\{ 3 \rho_i^2 \sin \left( \vartheta -2 \varphi_i
\right) -\rho_i^2 \sin \left( 3 \vartheta -2 \varphi_i \right)
+\right. \nonumber
\\%
& &\left. -\rho_i \rho_j \left[ 3 \sin \left( \vartheta -
\varphi_i -\varphi_j \right) -\sin \left( 3 \vartheta - \varphi_i
-\varphi_j \right) \right]\right\}%
\end{eqnarray}
\end{mathletters}

The first order perturbations only displace the caustic to the
centre of mass. Choosing this point as the origin of the
coordinate system, we can eliminate these terms. The second order
perturbations then yield the first non trivial deviations from the
Schwarzschild lens. These perturbations are in the form of
products between the masses forming the set of lenses, so there is
no superposition principle here.

This perturbative expansion is tightly related to the quadrupole
lens, as pointed out by Dominik (1999) for the binary lens. The
quadrupole lens equation has the form:
\begin{equation}
\mathbf{y} =\mathbf{x}-\frac{M \mathbf{x}}{\left|
\mathbf{x}\right|^2}+\frac{\left| \mathbf{x}\right|^2 \hat{Q}
\mathbf{x}-2\left(\mathbf{x}^\mathrm{T} \hat{Q} \mathbf{x} \right)
\mathbf{x}}{\left| \mathbf{x}\right|^6}
\end{equation}
where
\begin{equation}
\hat{Q}=\left( \begin{array}{cc}%
Q_1 & Q_2\\%
Q_2 & -Q_1%
\end{array} \right)
\label{Quadrupole moment}
\end{equation}
is the quadrupole moment.

A careful analysis of the second order Jacobian of the close
multiple system (\ref{Close Jacobian}) in the centre of mass
system reveals its coincidence with the Jacobian of the quadrupole
lens expanded to the first order in the quadrupole moment, in
polar coordinates. The elements of the matrix (\ref{Quadrupole
moment}) are:
\begin{equation}
\begin{array}{l}
Q_1=M \sum\limits_{i=1}^n m_i \rho_i^2 \cos \left( 2 \varphi_i
\right) \\%
Q_2=M \sum\limits_{i=1}^n m_i \rho_i^2 \sin \left( 2 \varphi_i
\right)
\end{array}
\label{Q elements}
\end{equation}

The two eigenvalues of the quadrupole matrix have the same
absolute value
\begin{equation}
Q=\sqrt{Q_1^2+Q_2^2} \label{Q eigenvalue}
\end{equation}
and opposite signs. We can define the orientation of the lens as
the direction of the eigenvector corresponding to the positive
eigenvalue, because it reduces to the positive $x_1$ direction in
the case of a binary system with the masses aligned on the
$x_1$--axis. The angle that this eigenvector forms with the
$x_1$--axis is:
\begin{equation}
\varphi=\arctan \frac{Q-Q_1}{Q_2}
\end{equation}

The caustic of a quadrupole lens is a diamond--shaped figure
symmetric for all reflections on the eigenvalues axes. If we
choose the coordinates so as to have the caustic oriented along
the $x_1$--axis, the four cusps are at the positions
$\vartheta=0,\frac{\pi}{2},\pi,\frac{3 \pi}{2}$. Fig. \ref{Fig
close binary caustics} shows some examples of central caustics for
a binary system with a mass ratio $q=3$. The second order
approximation works well up to separations of tenths of the total
Einstein radius. It is interesting to remark that the second order
result is always symmetric even if the system of lenses has no
symmetry at all. The symmetry is only lost when the perturbative
hypothesis begins to be more forced, as we see in Fig. \ref{Fig
close binary caustics}a.

\begin{figure}
 \resizebox{\hsize}{!}{\includegraphics{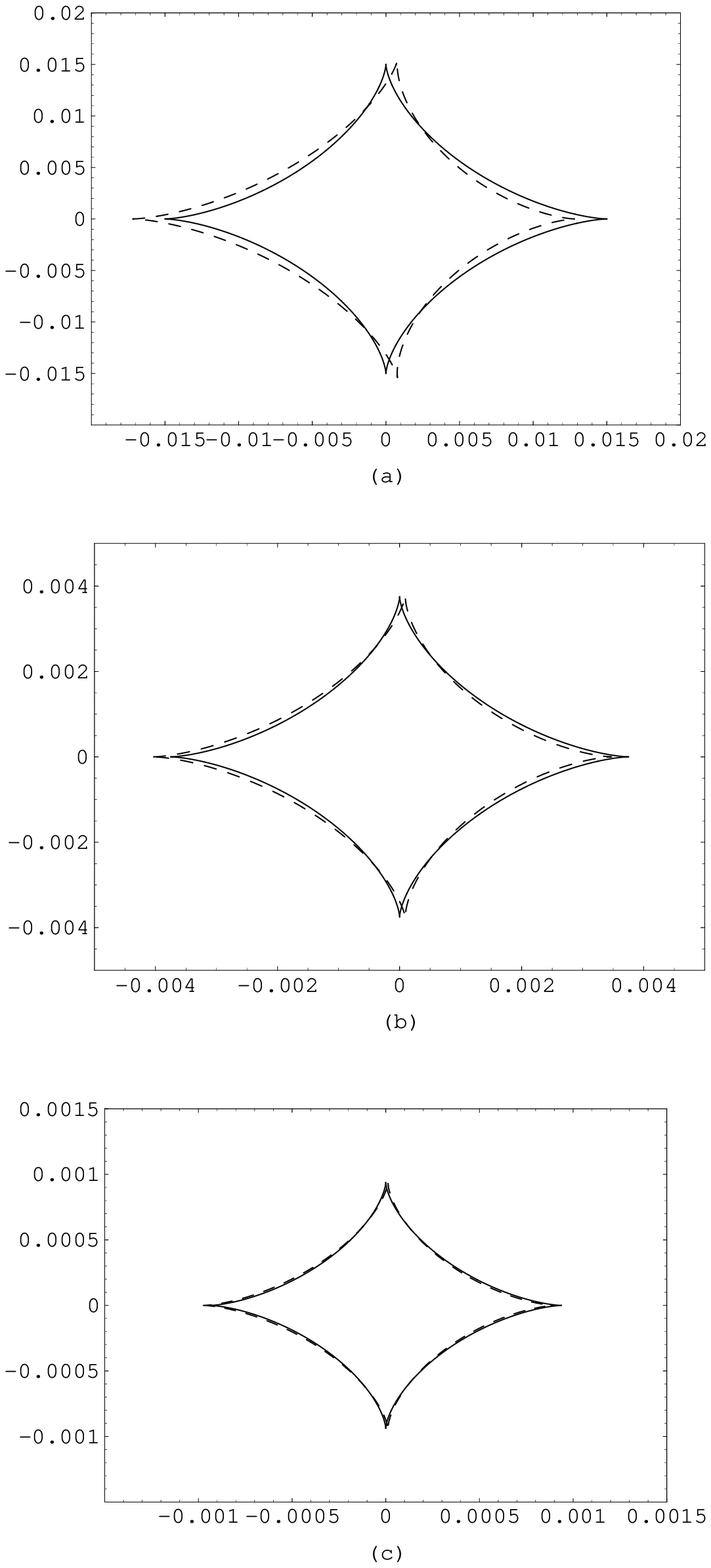}}
 \caption{Central caustic of a binary lens with $m_1=0.75$ and %
 $m_2=0.25$ and separation (a) 0.2, (b) 0.1, (c) 0.05. The dashed %
 curve is the exact caustic and the solid is the perturbative one.}
 \label{Fig close binary caustics}
\end{figure}

For the extension of the central caustic of a system of close
multiple lenses, we can directly refer to a well oriented
quadrupole lens caustic expanded to the first order in $Q$. The
area of such a caustic can be calculated with the same procedure
explained in the previous section for Chang--Refsdal caustics. The
result is:
\begin{equation}
A=\frac{3\pi Q^2}{2M^3}
\end{equation}

Recalling Eqs. (\ref{Q elements}) and (\ref{Q eigenvalue}), we see
that the area goes as the fourth power of the distances between
the masses, while the mass dependence is not so immediate because
of the presence of the cube of the total mass in the denominator.

\subsection{Secondary caustics}

\begin{figure}
 \resizebox{\hsize}{!}{\includegraphics{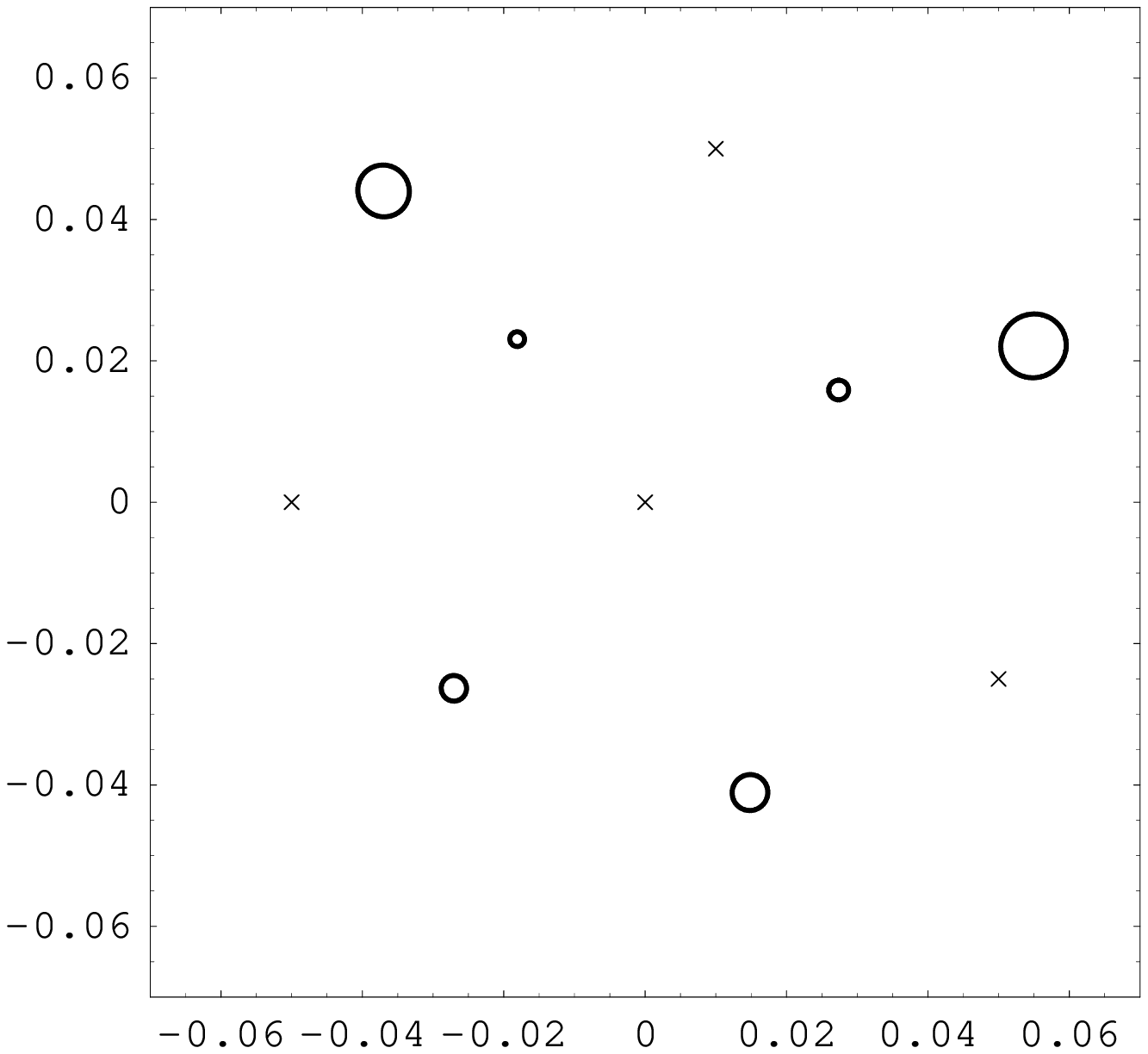}}
 \caption{Secondary critical curves generated by four %
 equal masses, represented by the small crosses in the figure.}
 \label{Fig close secondary critical curves}
\end{figure}

In the neighbourhood of the centre of mass of the system, some
critical curves in the form of small ovals can appear (see Fig.
\ref{Fig close secondary critical curves}). The number of these
curves is not fixed by the number of lenses but depends on the
position and the values of the masses composing the system. The
binary lens is the only case where the number of critical curves
is always fixed to two.

The perturbative approach can be used in the case of the binary
lens to find compact formulae for the positions of the secondary
critical curves and consequently the positions of the caustics.
The general multiple lens involves high degree equations which do
not allow any analytical considerations.

Let's consider two masses $m_1$ and $m_2$ separated by a distance
$a$ along the $x_1$--axis. In the centre of mass system, the
equation $\det J =0$ can be rationalized in the following form:
\begin{multline}
\left[ \left(x_1-\frac{a m_2}{M}\right)^2+x_2^2 \right]^2 \left[
\left(x_1+\frac{a m_1}{M}\right)^2+x_2^2 \right]^2+ \\%
-m_2^2 \left[ \left(x_1-\frac{a m_2}{M}\right)^2+x_2^2
\right]^2-m_1^2 \left[ \left(x_1+\frac{a m_1}{M}\right)^2+x_2^2
\right]^2+\\%
-2m_1 m_2 \left[ \left(x_1-\frac{a m_2}{M}\right)^2+x_2^2 \right]
\left[ \left(x_1+\frac{a m_1}{M}\right)^2+x_2^2 \right] +\\%
-8 m_1 m_2 x_2^2 \left(x_1-\frac{a m_2}{M}\right)
\left(x_1+\frac{a m_1}{M}\right)=0
\end{multline}
where $M=m_1+m_2$.

We search for solutions of the first order in the separation $a$.
So we put $x_1=a \Delta x_1$, $x_2=a \Delta x_2$ and expand in
powers of $a$. The first non trivial order is the fourth, so the
equation is of the fourth order in $\Delta x_1$ and $\Delta x_2$.
Introducing the polar coordinates $\Delta x_1= \Delta r \cos
\vartheta$, $\Delta x_2= \Delta r \sin \vartheta$, our equation
can be solved for $\Delta r$. The four solutions
\begin{multline}
\Delta r=\pm \frac{1}{M}\left[\left(m_2-m_1 \right) \cos \vartheta
+ \sqrt{m_1 m_2} \sin \vartheta + \right. \\%
\left. \pm i \left(\sqrt{m_1 m_2} \cos \vartheta +\left( m_1 -m_2
\right) \sin \vartheta \right) \right]%
\label{Delta r solutions}
\end{multline}
are real only if
\begin{equation}
\sqrt{m_1 m_2} \cos \vartheta +\left( m_1 -m_2 \right) \sin
\vartheta=0
\end{equation}
This equation is satisfied by the angles:
\begin{equation}
\vartheta=\pm \arccos \frac{\pm \left(m_2-m_1\right)}{
\sqrt{m_1^2+m_2^2-m_1 m_2}}
\end{equation}
Inserting these values in Eq. (\ref{Delta r solutions}),
discarding the negative solutions, and returning to the cartesian
coordinates, we have:
\begin{mathletters}
\begin{eqnarray}
\Delta x_1 & = & \pm \frac{m_1-m_2}{M} \\%
\Delta x_2 & = & \pm \frac{\sqrt{m_1 m_2}}{M}
\end{eqnarray}
\end{mathletters}
If we put these values in the original equation, there are only
two acceptable solutions:
\begin{mathletters}
\begin{eqnarray}
x_1 & = & - a \frac{m_1-m_2}{M} \\%
x_2 & = & \pm a \frac{\sqrt{m_1 m_2}}{M}
\end{eqnarray}
\end{mathletters}
These two points represent the centre of the small ovals at the
first order in the separation between the two masses. The shape of
the curves can be studied only at higher orders but high degree
polynomials heavily complicate the calculations.

Through the lens equation, we can find the positions of the
corresponding caustics:
\begin{mathletters}
\begin{eqnarray}
y_1 & = & \frac{m_1-m_2}{a} \\%
y_2 & = & \pm \frac{2\sqrt{m_1 m_2}}{a}
\end{eqnarray}
\end{mathletters}
We notice that when $a$ tends to zero, the two caustics become
infinitely far.

These formulae are compatible with those for the positions of the
couple of planetary caustics of planets internal to the Einstein
ring, given in (Bozza 1999), in the limit of planets very close to
the star.

\section{Planetary systems}

A particularly interesting case of multiple lensing is that of
planetary systems. Here a stellar mass is surrounded by planets
having masses much smaller. Their effects can usually be treated
as slight perturbations to the main lensing object.

The structure of the caustics of this lens has been widely
explored by perturbative methods in a previous work (Bozza 1999).
Here we shall recall the main formulae to complete the picture of
perturbative results in multiple lensing and make some additional
considerations.

We can distinguish between the central caustic which is generated
by the distortion of the star's Einstein ring and the planetary
caustics which are the images of the planetary critical curves and
can be well approximated by Chang--Refsdal critical curves.

\subsection{Central caustic}

The central caustic of a star with mass $m_1$ placed at the
origin, surrounded by planets with masses $m_2, ..., m_n$ placed
at $\mathbf{x}_2, ..., \mathbf{x}_n$ respectively is expressed by
the following parametric form:
\begin{mathletters}
\label{Central caustic}
\begin{eqnarray}
y_{1}\left( \vartheta \right) & =2\sqrt{m_{1}}\varepsilon \left(
\vartheta
\right) \cos \vartheta -\sum\limits_{i=2}^{n}\frac{m_{i}\Delta _{i1}^{0}}{%
\left[ \left( \Delta _{i1}^{0}\right) ^{2}+\left( \Delta
_{i2}^{0}\right) ^{2}\right] } \\ y_{2}\left( \vartheta \right) &
=2\sqrt{m_{1}}\varepsilon \left( \vartheta
\right) \sin \vartheta -\sum\limits_{i=2}^{n}\frac{m_{i}\Delta _{i2}^{0}}{%
\left[ \left( \Delta _{i1}^{0}\right) ^{2}+\left( \Delta
_{i2}^{0}\right) ^{2}\right] }
\end{eqnarray}
\end{mathletters}
where
\begin{multline}
\varepsilon \left( \vartheta \right) =\frac{1}{2} \cos 2\vartheta
 \sum\limits_{i=2}^{n}m_{i}
\frac{\left(
\Delta _{i1}^{0}\right) ^{2}-\left( \Delta _{i2}^{0}\right) ^{2} }{%
\left[ \left( \Delta _{i1}^{0}\right) ^{2}+\left( \Delta
_{i2}^{0}\right) ^{2}\right] ^{2}}+  \label{Central critic curve}
\\ +\sin 2\vartheta
\sum\limits_{i=2}^{n}m_{i}\frac{\Delta _{i1}^{0}\Delta
_{i2}^{0}}{\left[ \left( \Delta _{i1}^{0}\right) ^{2}+\left(
\Delta _{i2}^{0}\right) ^{2}\right] ^{2}}
\end{multline}
and $\mathbf{\Delta }_{i}^0 = \left( \sqrt{m_1} \cos
\vartheta-x_{i1}; \sqrt{m_1} \sin \vartheta-x_{i2} \right)$.

In many studies of the central caustic of a planetary system, a
principle of duality between planets external and internal to the
Einstein ring (hereafter, we shall simply refer to them as
external and internal planets, respectively) was often claimed on
the basis of the observation of the shape of numerical caustics
(Griest \& Safizadeh 1997; Dominik 1999). This principle can be
directly verified on these formulae which are invariant under the
transformation
\begin{equation}
\mathbf{x}_i \rightarrow \frac{m_1}{\left| \mathbf{x}_i \right|^2}
\mathbf{x}_i
\end{equation}
So the conjecture of duality has an effective analytical basis.

The central caustic is generally a self--intersecting curve,
except for the case of the single planet, when it has the already
encountered diamond shape. Then, its area can be calculated with
the same method of the previous sections. The result is:
\begin{equation}
A=\frac{\pi m_2^2}{4m_1}\left( 1-\frac{ \left( m_1 + \rho_2^2
\right) \left(m_1^2 -4m_1 \rho_2^2+ \rho_2^4 \right) }{\left|
m_1-\rho_2^2 \right|^3}
\right)%
\label{Central caustic area}
\end{equation}

It positively diverges when $\rho_2 \rightarrow \sqrt{m_1}$. This
happens because in this limit the perturbative approach is no
longer valid and the fusion between the central and the planetary
caustic occurs. The area vanishes when $\rho_2 \rightarrow 0$ or
$\rho_2 \rightarrow \infty$ because, in these limits, the
Schwarzschild lens is recovered and the central caustic reduces to
a point.

\subsection{Planetary caustics}

The planetary critical curve is always localized in the
neighbourhood of the planet and assumes the shape of an elongated
ring, when the planet is outside of the star's Einstein ring, or
splits into two specular ovals, when the planet is inside.

The zero point of the expansion of these critical curves is then
the position $\mathbf{x}_2$ of the planet we are considering. The
first non trivial order is $\mathbf{x}-\mathbf{x}_2 \sim
\sqrt{m_2}$, since the critical curve of a very far planet is
nothing but its Einstein ring with radius $\sqrt{m_2}$.

In (Bozza 1999), the first order planetary critical curves were
derived and their relations with the Chang--Refsdal ones was
discussed. It is interesting to go farther in the expansion to
understand how this relation is broken and to see the effects of
other planets.

It is useful to consider polar coordinates around the planet
position $\mathbf{x}_2$. Let's then set $\mathbf{x} - \mathbf{x}_2
= \left(r \cos \vartheta; r \sin \vartheta \right)$. Now we expand
the radial coordinate $r$:
\begin{equation}
r=\varepsilon_1+\varepsilon_2+...
\end{equation}
where $\varepsilon_i \sim m_2^{i/2}$.

The first two orders of the equation $\det J =0$ give:
\begin{multline}
1-\frac{m_1^2}{\rho_2^4}-\frac{m_2^2}{\varepsilon_1^4} - \frac{2
m_1 m_2 \cos \left( 2\vartheta-2 \varphi_2
\right)}{\varepsilon_1^2 \rho_2^2}+ \label{Planetary EqdetJ}\\%
+\frac{4 m_2^2}{\varepsilon_1^5} \varepsilon_2+\frac{4 m_1 m_2
\cos \left( 2\vartheta-2 \varphi_2 \right)}{\varepsilon_1^3
\rho_2^2} \varepsilon_2+\\%
+\frac{4 m_1 \left[m_1 \varepsilon_1^2 \cos \left( \vartheta-
\varphi_2 \right) + m_2 \rho_2^2 \cos \left( 3\vartheta- 3
\varphi_2 \right) \right]}{\varepsilon_1 \rho_2^5}=0
\end{multline}

We see that the first row, representing the lowest order Jacobian,
is just the Jacobian of the Chang--Refsdal lens $m_2$ with shear
$\gamma=\frac{m_1}{\rho_2^2}$. Then the lowest order critical
curve is exactly Chang--Refsdal:
\begin{equation}
\varepsilon_1=\sqrt{m_2 \frac{\frac{m_1}{\rho_2^2} \cos \left(
2\vartheta-2 \varphi_2 \right) \pm \sqrt{1- \frac{m_1^2}{\rho_2^4}
\sin^2 \left( 2\vartheta-2 \varphi_2 \right)}
}{1-\frac{m_1^2}{\rho_2^4}}}%
\label{Planetary critical curve}
\end{equation}
According to the double sign, two branches are present. For
external planets only the higher is real, while for internal
planets both branches are real in two intervals centred on
$\vartheta=\pm \frac{\pi}{2}$ (Bozza 1999).

We can notice that if we perform an expansion of Eq.
(\ref{Planetary critical curve}) to the first order in the shear,
we obviously get the same result of Sect. 3, with the roles of
$m_2$ and $m_1$ interchanged. However, in this section we are
analysing the perturbations in the masses of the planets, so this
expansion is not interesting for our purposes, though it gives the
connection with the previous calculation.

The full Eq. (\ref{Planetary EqdetJ}) can now be employed to find
the second order perturbation:
\begin{equation}
\varepsilon_2=-\frac{ m_1 \varepsilon_1^4 \left[m_1
\varepsilon_1^2 \cos \left( \vartheta- \varphi_2 \right) + m_2
\rho_2^2 \cos \left( 3\vartheta- 3 \varphi_2 \right) \right]}{m_2
\rho_2^3 \left[m_1 \varepsilon_1^2 \cos \left( 2 \vartheta-
2\varphi_2 \right) + m_2 \rho_2^2 \right]}
\end{equation}

As usual, through the lens equation we can write down the formulae
for the caustics:
\begin{mathletters}
\begin{eqnarray}
y_1\left( \vartheta \right) = & \left( \rho_2 -\frac{m_1}{\rho_2}
\right) \cos \varphi_2 + \left( \varepsilon_1
-\frac{m_2}{\varepsilon_1} \right) \cos
\vartheta + \nonumber \\%
&+ \frac{m_1 \varepsilon_1 \cos \left( \vartheta - 2 \varphi_2
\right) }{\rho_2^2}+ \left( 1+ \frac{m_2}{\varepsilon_1^2} \right)
\varepsilon_2 \cos \vartheta + \nonumber \\%
& +m_1\frac{  \varepsilon_2 \rho_2 \cos \left( \vartheta -2
\varphi_2 \right) -\varepsilon_1^2 \cos \left( 2\vartheta -3
\varphi_2 \right)  }{\rho_2^3} + \nonumber \\%
& - \sum\limits_{i=3}^n m_i \frac{\rho_2 \cos \varphi_2 - \rho_i
\cos \varphi_i }{\rho_2^2 + \rho_i^2 -2 \rho_2 \rho_ i \cos \left(
\varphi_2 - \varphi_i \right) }\\%
y_2\left( \vartheta \right)= & \left( \rho_2 -\frac{m_1}{\rho_2}
\right) \sin \varphi_2 + \left( \varepsilon_1
-\frac{m_2}{\varepsilon_1} \right) \sin
\vartheta + \nonumber \\%
& -\frac{m_1 \varepsilon_1 \sin \left( \vartheta - 2 \varphi_2
\right) }{\rho_2^2}+ \left( 1+ \frac{m_2}{\varepsilon_1^2} \right)
\varepsilon_2 \sin \vartheta + \nonumber \\%
& +m_1\frac{  \varepsilon_2 \rho_2 \sin \left( \vartheta -2
\varphi_2 \right) -\varepsilon_1^2 \sin \left( 2\vartheta -3
\varphi_2 \right) }{\rho_2^3} + \nonumber \\%
& - \sum\limits_{i=3}^n m_i \frac{\rho_2 \sin \varphi_2 - \rho_i
\sin \varphi_i }{\rho_2^2 + \rho_i^2 -2 \rho_2 \rho_i \cos \left(
\varphi_2 - \varphi_i \right) }
\end{eqnarray}
\end{mathletters}

At the second order a sum containing the effects of the other
planets appears. It does not depend on $\vartheta$ and then only
represents a displacement of the caustic towards the other
planets.

\begin{figure}
 \resizebox{\hsize}{!}{\includegraphics{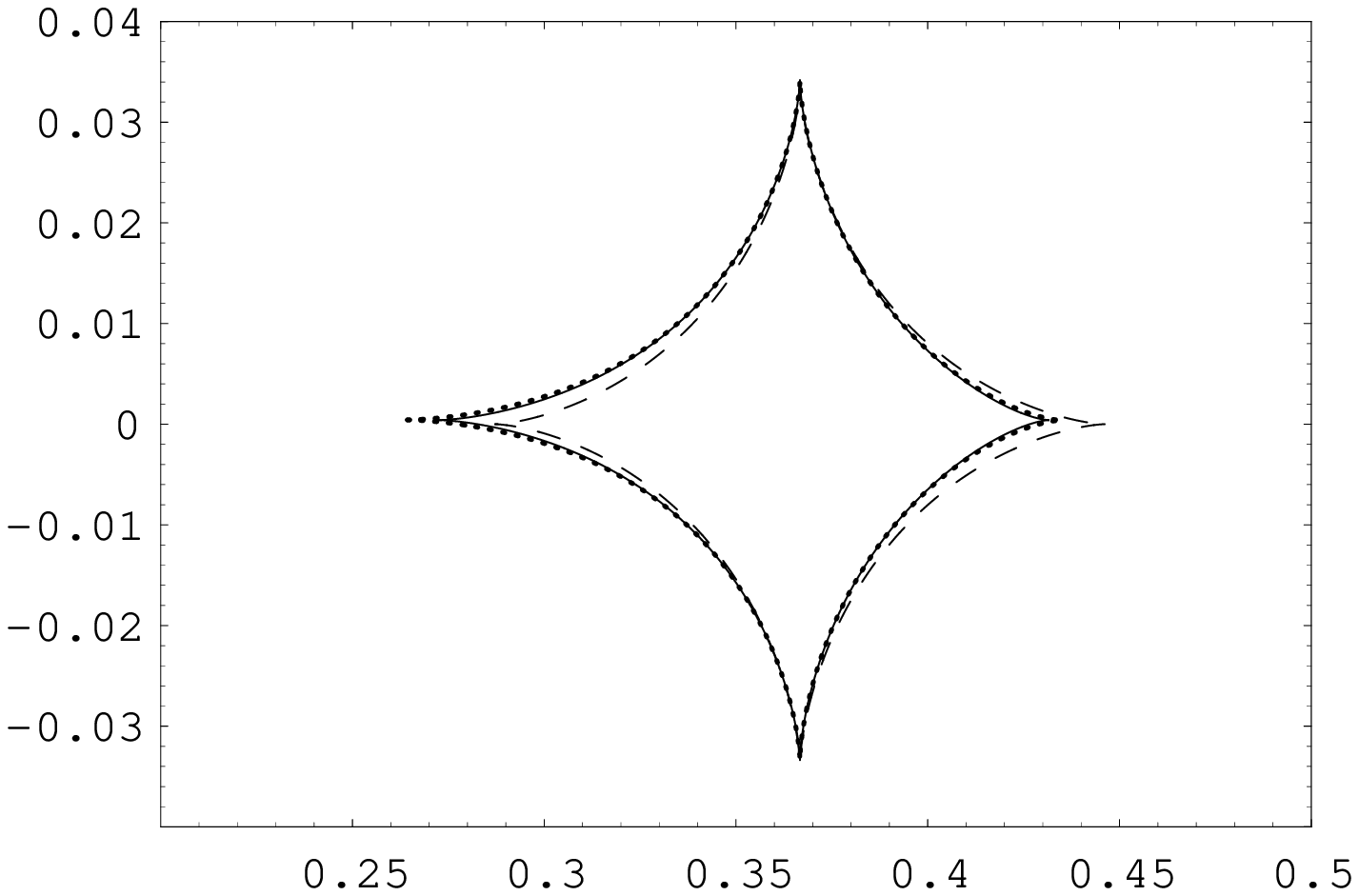}}
 \caption{Planetary caustic of a jovian planet ($m_2=10^{-3}$) %
 at position $\left( 1.2; 0 \right)$. The caustic is perturbed by %
 the presence of another jovian planet at $\left( 0; 1.5 \right)$. %
 The dashed curve is the first order approximation, the solid one is %
 the second order approximation, which is almost coincident with the %
 exact one (dotted curve).}
 \label{Fig planetary caustic}
\end{figure}

Fig. \ref{Fig planetary caustic} shows a comparison between the
perturbative results and the exact curve for an external planet.
The loss of symmetry of the figure, not present in the first
approximation, is reliably followed by the second order curve.
Moreover, it is to notice that the displacement of the caustic,
due to other planets, is correctly reproduced.

The critical curves of internal planets are not defined for all
values of $\vartheta$. This fact brings in divergences at the
border points where the solution switch from imaginary to real. So
the second order results are very good everywhere except for a
neighbourhood of these border values of the angle.

As for the other caustics, we can easily calculate the area of
planetary caustics exploiting their symmetries. The case of
external planets is analogue to the previous ones, so we can
directly give the result:
\begin{equation}
A=2m_2 \left[ \pi -4 E \left( \frac{\pi}{2} ;
\frac{m_1^2}{\rho_2^4} \right)+2 F\left(\frac{\pi}{2} ;
\frac{m_1^2}{\rho_2^4} \right) \right]
\end{equation}
where:
\begin{eqnarray}
F \left(\varphi ;  m \right) & = &\int\limits_0^{\varphi} \left[
1-m
\sin^2 \vartheta \right]^{-1/2} d \vartheta \\%
E \left(\varphi ;  m \right)& = & \int\limits_0^{\varphi} \left[
1-m \sin^2 \vartheta \right]^{1/2} d \vartheta
\end{eqnarray}
are the elliptic integrals of the first and the second kind
respectively.

The calculation of the area for the caustics of the internal
planets is not very different. We have to integrate on the higher
branch and subtract the integral of the lower so as to obtain the
area included between the two branches. The formula (\ref{Area
formula}) must then be modified this way:
\begin{equation}
A=\int\limits_{\vartheta_{\mathrm{min}}}^{\vartheta_{\mathrm{max}}}
\left(y_2\left( \vartheta \right)
\frac{\mathrm{d}y_1}{\mathrm{d}\vartheta} \right)_{h.b.}
\mathrm{d}\vartheta
-\int\limits_{\vartheta_{\mathrm{min}}}^{\vartheta_{\mathrm{max}}}
\left(y_2\left( \vartheta \right)
\frac{\mathrm{d}y_1}{\mathrm{d}\vartheta} \right)_{l.b.} \mathrm{d}\vartheta%
\end{equation}
where $\vartheta_{\mathrm{min}}$ and $\vartheta_{\mathrm{max}}$
are the extremes of the interval where the two branches are real.

The result is:
\begin{multline}
A=m_2 \left( 2 E \left( 2 \vartheta_{\mathrm{min}};
\frac{m_1^2}{\rho_2^4}
\right) -2 E \left(  2 \vartheta_{\mathrm{max}}; \frac{m_1^2}{\rho_2^4} \right)+ \right. \\%
\left.- F \left( 2 \vartheta_{\mathrm{min}};
\frac{m_1^2}{\rho_2^4} \right) + F \left( 2
\vartheta_{\mathrm{max}}; \frac{m_1^2}{\rho_2^4} \right) \right)
\end{multline}
Of course, as there are two planetary caustics for internal
planets, this number must be doubled.

\begin{figure}
 \resizebox{\hsize}{!}{\includegraphics{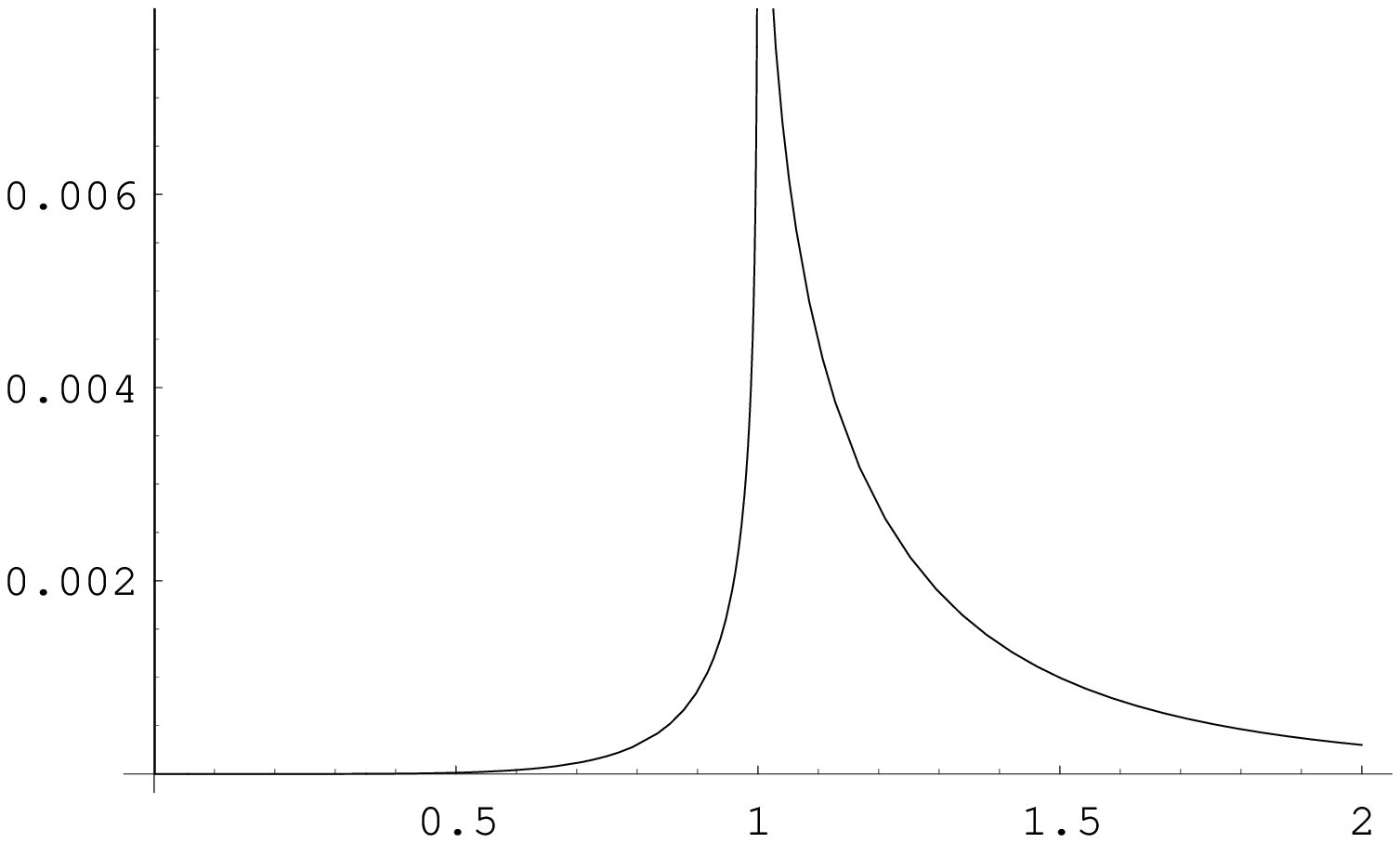}}
 \caption{Area of the planetary caustics as a function of %
 the distance of the planet from the star.}
 \label{Fig planetary area}
\end{figure}

The area of the planetary caustics is of order $m_2$. In fact it
is well known that they are much greater than the central caustic,
which is instead of order $m_2^2$ (see Eq. (\ref{Central caustic
area})) (Dominik 1999).

In Fig. \ref{Fig planetary area} we see a plot of the area of the
planetary caustic as a function of the separation of the planet
from the star. We see the divergence in correspondence of the
``resonant lensing'' situation ($\rho_2=\sqrt{m_1}$) where the
perturbative theory cannot be applied.

\section{Conclusions}

The results presented in this work prove the great flexibility of
perturbative methods in the resolution of problems in
gravitational lensing. By different choices of perturbative
expansions, it is possible to obtain very useful analytical
approximations, which can be employed for further applications. In
the case of a discrete set of point-like lenses, we gain a deep
understanding of the distortion effects with respect to the
Schwarzschild lens. This is the only analytical knowledge we have
about the shape of caustics in multiple lenses.

Carrying the expansions to higher orders allows some interesting
considerations about the validity of some approximations such as
the Chang--Refsdal or the quadrupole lens and the limits of
applicability of a superposition principle for the effects coming
from different masses.

The calculation of the area of the caustics is another important
possibility offered by the achievement of these analytical
approximations.

\begin{acknowledgements}
I would like to thank Gaetano Lambiase for his contributions to
the birth of the ideas behind this work.
\end{acknowledgements}

\end{document}